\documentclass[graybox,natbib,nosecnum]{svmult}
\bibpunct{(}{)}{;}{a}{}{,} 

\pdfoutput=1   
\usepackage{mathptmx}           
\usepackage{helvet}                 
\usepackage{courier}                
\usepackage{type1cm}              
\usepackage{makeidx}               
\usepackage{graphicx}              
\usepackage{multicol}                
\usepackage[bottom]{footmisc} 
\usepackage[normalem]{ulem}	
\usepackage{hyperref}              
\usepackage{soul}                     
\usepackage{amsmath}	        
\usepackage{amssymb}	         

\newcommand*\aap{A\&A~}

\newcommand*\aj{AJ~}

\newcommand*\apj{ApJ~}
\newcommand*\apjl{ApJ~}

\newcommand*\apjs{ApJS~}

\newcommand*\grl{Geophys Res Lett~}

\newcommand*\icarus{Icarus~}

\newcommand*\jgr{J Geophys Res~}
\newcommand*\jqsrt{J Quant Spectr Rad Transf~}

\newcommand*\mnras{MNRAS~}

\newcommand*\nat{Nature~}

\newcommand*\pasp{PASP~}

\newcommand*\planss{Planet Space Sci~}

\newcommand*\procspie{Proc SPIE~}
\newcommand*\psj{Planet Sci J~}

\makeindex             

\begin{document}

\title*{Characterizing Exoplanets for Habitability}
\author{Tyler D. Robinson}
\institute{Tyler D. Robinson \at Lunar and Planetary Laboratory, University of Arizona, Tucson, AZ 85721, USA, \email{tdrobin@arizona.edu}}
%
%
\maketitle

\abstract{A habitable exoplanet is a world that can maintain stable liquid water on its surface.  Techniques 
and approaches to characterizing such worlds are essential, as performing a census of Earth-like planets 
that may or may not have life will inform our understanding of how frequently life originates and is sustained 
on worlds other than our own.  Observational techniques like high contrast imaging and transit spectroscopy 
can reveal key indicators of habitability for exoplanets.  Both polarization measurements and specular 
reflectance from oceans (also known as ``glint'') can provide direct evidence for surface liquid water, while 
constraining surface pressure and temperature (from moderate resolution spectra) can indicate liquid 
water stability.  Observations of variability (that indicates weather) from, as well as mapping of, exoplanets 
can provide indirect evidence of habitability, and measurements of water vapor or cloud profiles that 
indicate condensation near a surface could also provide evidence for habitability.  Approaches to making 
the types of measurements that indicate habitability are diverse, and have different considerations 
for the required wavelength range, spectral resolution, maximum noise levels, stellar host temperature, 
and observing geometry.}

\section{Introduction }

The emphasis on which planetary surface properties are key to determining habitability has, 
like many ideas in exoplanetary science, changed in time.  Early work, even before the first 
detections of worlds around other stars, highlighted the importance of clement conditions at the 
planetary surface---\citet{huang1959} discussed the need for enough ``heat'' for organisms to 
survive, while \citet{dole1964} focused on a temperature range that would be suitable for humans.  
However, it has become widely accepted that liquid water played an essential role in the origin, 
development, and maintenance of life on Earth \citep[e.g.,][]{brack1993}, and, as a result, most 
studies of habitability have since focused on stable surface liquid water.  \citet{rasool&debergh1970}, 
in their early description of the runaway greenhouse and the evolution of climate on the terrestrial 
planets of the Solar System, were amongst the first to place a strong emphasis on stable surface 
liquid water.  Similarly, in their study of the co-evolution of Earth and life, \citet{hart1978} required 
liquid water (as well as certain atmospheric composition constraints) for life to originate.  

Foundational work on habitable zones around other stars 
\citep{hart1979,whitemireetal1991,kastingetal1993} and on the emergence of habitable worlds 
\citep{matsui&abe1986,abe&matsui1988} strongly emphasize that habitable planets are the subset 
of terrestrial worlds that can have stable surface liquid water.  Of course, this does not rule out the 
potential for sub-surface habitable environments, such as the ocean beneath Europa's icy crust 
\citep{pappalardoetal1999,chyba&phillips2001}---the emphasis on {\it surface} habitability is based 
in pragmatism, as discovering and characterizing any sub-surface habitable environments on 
exoplanets (or exomoons) will likely remain unfeasible even into the distant future.  

Motivated in large part by the success of NASA's {\it Kepler} mission \citep{boruckietal2010} and the 
ever-growing number of known potentially-habitable exoplanets, many authors and groups have 
sought to refine our understanding of the myriad ways a planet can (or cannot) maintain habitability 
\citep{forget&pierrehumbert1997,joshietal1997,stevenson1999,selsisetal2007,haqqmisraetal2008,
vonparisetal2010,abeetal2011,pierrehumbert&gaidos2011,goldblattetal2013,kopparapuetal2013,
rugheimeretal2013,wordsworth&pierrehumbert2013,wolf&toon2013,leconteetal2013,yangetal2013,
shieldsetal2013,zsometal2013,ramirez&kaltenegger2014,luger&barnes2015,wayetal2016}.
These studies demonstrate a relatively well-developed understanding of {\it how} a world can be 
habitable.  The next critical step towards undertaking a census of Earth-like planets around our 
nearest stellar neighbors is, then, developing techniques for remotely assessing the likelihood that 
an exoplanet could have stable surface liquid water or, in other words, the remote characterization 
of habitability.

Interestingly, Galileo was likely the first to consider the remote detection of surface liquid water 
when, in his {\it Dialogue Concerning the Two Chief World Systems}, he used the brightness of 
the dark portion of the lunar disk (which is illuminated by reflected light from Earth) to deduce 
that ``seas would appear darker, and [\ldots] land brighter'' when observed from a distance 
\citep{galileo1632}.  Modern discussions of techniques for the remote characterization of 
exoplanet habitability, which are the focus of this chapter, generally fall into three categories.  
First, surface liquid water could be directly detected using reflection and/or polarization 
measurements.  Second, the surface pressure and temperature of an exoplanet could be inferred 
from spectroscopic observations, which then determines the stability of liquid water from its phase 
diagram.  Finally, in the absence of such direct evidence for surface liquid water (or its stability), 
habitability would need to be constrained using some combination of photometric and 
spectroscopic observations and modeling to try to best understand the planetary surface 
environment.

The direct detection of surface liquid water and the inference of surface pressure and 
temperature were demonstrated for Earth by \citet{saganetal1993} using observations 
from a pair of flybys performed by the {\it Galileo} spacecraft.  Specular reflection, which is 
indicative of liquids, was detected with {\it Galileo} imaging, and solid phase water absorption 
was identified in observations of the polar caps, thereby providing strong evidence that the 
surface liquid is water.  Near-infrared spectral observations of thermal emission from 
cloud-free regions of the planet revealed surface temperatures in the range 
240--290~K, and a crude spectral retrieval analysis \citep{drossartetal1993} indicated an 
integrated atmospheric column mass $>\!200$~g~cm$^{-2}$.  For an Earth-like gravitational 
acceleration, this column mass corresponds to a surface pressure $>\!0.2$~bar.  Given the 
aforementioned range of surface temperatures, the surface (even for pressures much 
larger than the lower limit quoted here) spans the solid-to-liquid transition regime for water, 
and is, thus, habitable.

\citet{saganetal1993} also present multiple strong lines of evidence for Earth being inhabited.  
This indicates that, for some worlds, it may prove easier to detect life than to detect habitability.  
It is, however, critical that approaches be developed which can be used to independently recognize 
both habitable as well as life-bearing planets.  Employing these independent approaches, as part 
of a larger census of exoplanetary surface and atmospheric environments, will tell us if originating 
and sustaining life is common  (where nearly all potentially habitable worlds are inhabited) or rare 
(where nearly all planets that show signs of habitability do not show signs of life).  Either of these 
findings would tell us something profound about our place in the Universe.

From the perspective of exoplanet science, the \citet{saganetal1993} {\it Galileo} results 
are missing a key complication---the habitability analyses all rely on spatially resolved 
observations.  Generally, observations of exoplanets are spatially unresolved.  Thus, for 
a true Pale Blue Dot, cloudy and clearsky, ocean and land, and warm and cold scenes would 
all be blended together, which significantly complicates our ability to characterize the surface 
environment for signs of habitability.  Here, it must be emphasized that habitability is a {\it surface}  
phenomenon and can only be constrained if a remote observation has sensitivity to the 
surface (i.e., that some light at certain wavelengths in the observed spectral range comes 
from at/near the surface).  In other words, we would have little hope of studying the surface 
environment of a terrestrial exoplanet that is enshrouded with completely opaque clouds.

The following sections present and synthesize studies related to the characterization of 
exoplanet habitability.  For earlier reviews on characterizing terrestrial exoplanets which 
include some details on habitability, see \citet{meadows2010} \citet{kalteneggeretal2010}.  A 
complementary review of processes related to planetary habitability and its remote characterization
can be found in \citet{kopparapuetal2020}. In our review, we begin 
with an overview of the key observational techniques that can be used to remotely characterize 
exoplanets, and highlight the sizes of signatures relevant to studying Pale Blue Dots.  Following 
this overview, we discuss how the different observational techniques can be used to directly 
detect surface liquid water, to measure surface pressure and temperature, and/or to place other 
key constraints on the planetary environment.  Whenever possible, the feasibility of detecting 
habitability indicators is discussed.  We conclude by outlining several important questions that 
remain unaddressed on the topic of characterizing for habitability.

\section{Observational Techniques}

Several observational techniques are relevant to the characterization of the atmospheres 
and surface environments of potentially habitable exoplanets: transit spectroscopy, high 
contrast imaging, and secondary eclipse spectroscopy.  We briefly review these here and  
demonstrate the relevant signal sizes.  For an overview of techniques and signature sizes 
for a diversity of planet types, see \citet{cowanetal2015}.

\runinhead{Transit Spectroscopy:}  In transit spectroscopy 
\citep{seager&sasselov2000,brown2001,hubbardetal2001}, the small fractional 
dimming of an unresolved exoplanet host star is measured as the planet 
transits the stellar disk.  This quantity---the transit depth---is usually interpreted as the 
square of the ratio of a characteristic planetary radius ($R_{\rm{p}}$) to the stellar radius 
($R_{\rm{s}}$), and, when measured at different wavelengths, the transit depth indicates 
the planetary atmospheric opacity as the world will appear larger on the stellar disk at 
wavelengths that correspond to larger extinction.  While the overall scale of the transit 
depth is given by $\left( R_{\rm{p}}/R_{\rm{s}} \right)^{\!2}$, the contrast of spectral features 
will depend on the altitude difference probed within versus outside a molecular band 
($\Delta z$), and is approximately, 
\begin{equation}
  \frac{2\Delta z R_{\rm{p}}}{R_{\rm{s}}^2} \approx 0.6~{\rm ppm} ~ \left(\frac{T}{250~{\rm{K}}}\right)  
  \left(\frac{29~{\rm{g}}~{\rm{mol}^{-1}} }{\mu}\right)  \left(\frac{5.5~{\rm{g}}~{\rm{cm}^{-3}} }{\rho_{\rm{p}}}\right)  
  \left(\frac{R_{\odot}}{R_{\rm{s}}}\right)^{\!\!2} \ ,
\end{equation}
where $T$ is a characteristic atmospheric temperature, $\mu$ is the atmospheric mean 
molar weight, and $\rho_{\rm{p}}$ is the planetary bulk density.  We have assumed the 
altitude range probed is a few pressure scale heights, and we have adopted Earth-like 
values for all parameters.  For early, mid, and late M dwarfs, the scale of features 
increases to 2, 10, and 60~ppm, respectively.

\runinhead{High Contrast Imaging:}  In high contrast (or ``direct'') imaging 
\citep{traub&oppenheimer2010}, optical 
techniques are used to resolve the faint point spread function of a planetary companion 
from that of its bright host.  Typical approaches include coronagraphy 
\citep{guyonetal2006,mawetetal2012}, external occulters or ``starshades'' 
\citep{cashetal2007,shaklanetal2010}, and interferometry \citep{beichmanetal1999}.  The 
relevant measure is the planet-to-star flux ratio ($F_{\rm{p}}/F_{\rm{s}}$), which (roughly) 
sets the contrast that must be achieved to accomplish imaging (although planet-star 
angular separation, host star apparent magnitude, exozodiacal dust brightness, and 
other quantities also impact the feasibility of observation).  For reflected light, which 
would be the focus of any near- or far-future direct imaging efforts, the flux ratio is given 
by $A_{\rm{g}} \Phi (\alpha) (R_{\rm{p}}/a)^2$, where $A_{\rm{g}}$ is the geometric 
albedo, $\Phi$ is the phase function (which depends on the phase angle, $\alpha$), 
and $a$ is the orbital distance.  Assuming that the insolation on potentially habitable 
exoplanets is roughly that of what Earth receives ($S_{\oplus}=1360$~W~m$^{-2}$), 
we have,
\begin{equation}
  \frac{F_{\rm{p}}}{F_{\rm{s}}} \approx 1\!\times\!10^{-10} ~ \left(\frac{A_{\rm{g}}}{0.2}\right)  
  \left(\frac{\Phi \! \left( 90^{\circ} \right)}{\pi}\right)  \left(\frac{S_{\rm{p}}}{S_{\oplus}}\right) 
  \left(\frac{R_{\rm{p}}}{R_{\oplus}}\right)^{\!\!2} \left(\frac{R_{\odot}}{R_{\rm{s}}}\right)^{\!\!2}  
  \left(\frac{T_{{\rm{eff,}}\odot}}{T_{{\rm{eff,s}}}}\right)^{\!\!4} \ ,
\end{equation}
where $T_{\rm{eff}}$ is a stellar effective temperature, and values for an Earth-like 
V-band geometric albedo and phase function (at quadrature) are adopted.  Note that 
it is important to distinguish between planet detection (which is driven by the 
planet-to-star flux ratio) and atmospheric characterization.  The latter requires 
detecting spectral features (and, possibly, the base of these features), which can be 
at substantially smaller planet-to-star flux ratios and, owing to the overall faintness of 
the planet in these features, may drive long integration times.  For early, 
mid, and late M dwarfs, the flux ratio increases dramatically to $2\!\times\!10^{-9}$, 
$5\!\times\!10^{-8}$, and $4\!\times\!10^{-7}$, respectively.  For these cooler stars, 
though, the small inner working angle that would be needed to resolve the planet 
from the star drives the need for large-diameter telescopes.  Adopting 
$2\lambda/D$ as a ``practical'' limit to small inner working angle photometry 
\citep{mawetetal2014}, we see that the telescope diameter required to resolve a 
habitable zone planet from its host is roughly,
\begin{equation}
  D > 4 ~ {\rm{m}} ~ \left( \frac{\lambda}{1 ~ \mu{\rm{m}}} \right) 
  \left( \frac{d}{10 ~ {\rm{pc}}} \right) \left(\frac{S_{\rm{p}}}{S_{\oplus}}\right)^{\!\!1/2} 
  \left(\frac{R_{\odot}}{R_{\rm{s}}}\right) \left(\frac{T_{{\rm{eff,}}\odot}}{T_{{\rm{eff,s}}}}\right)^{\!\!2} \ ,
\end{equation}
where $\lambda$ is wavelength, $D$ is the telescope diameter, and $d$ is the 
distance to the system.  Returning to the early, mid, and late M dwarf cases, the 
lower limits on the diameter are 16, 90, and 250~m, respectively.  

\runinhead{Secondary Eclipse Spectroscopy:}  Like transit spectroscopy, 
secondary eclipse spectroscopy is a differential measurement that requires the 
combined planetary and stellar flux prior to the planet disappearing behind its 
host star and comparing this to the stellar flux measured during eclipse 
\citep{winn2010}.  Here, as was the case for direct imaging, the key quantity is 
the planet-to-star flux ratio (at full phase).  Taking a mid-M dwarf as an example, 
in reflected light we have,
\begin{equation}
  \frac{F_{\rm{p}}}{F_{\rm{s}}} \approx 0.2~{\rm ppm} ~ \left(\frac{A_{\rm{g}}}{0.2}\right)  
  \left(\frac{S_{\rm{p}}}{S_{\oplus}}\right) 
  \left(\frac{R_{\rm{p}}}{R_{\oplus}}\right)^{\!\!2} \left(\frac{0.2 \cdot R_{\odot}}{R_{\rm{s}}}\right)^{\!\!2}  
  \left(\frac{2800~{\rm{K}}}{T_{{\rm{eff,s}}}}\right)^{\!\!4} \ ,
\end{equation}
which is quite small.  The characteristic signature size improves at thermal 
wavelengths, as the planet is self-luminous at these wavelengths.  Here, we 
have the ratio of two blackbodies, and taking the stellar spectrum to be in the 
Rayleigh-Jeans limit, we have,
\begin{equation}
  \frac{F_{\rm{p}}}{F_{\rm{s}}} \approx 3~{\rm ppm} \left(\frac{R_{\rm{p}}}{R_{\oplus}}\right)^{\!\!2} 
  \left(\frac{0.2 \cdot R_{\odot}}{R_{\rm{s}}}\right)^{\!\!2} \left( \frac{2800 ~ {\rm{K}}}{T_{{\rm{eff,s}}}} \right) 
  \left( \frac{\lambda}{10 ~ \mu{\rm{m}}} \right)^{\!\!4} \!
  \left( \frac{B_{\lambda}(T)}{B_{10~\!\mu{\rm{m}}}(250~{\rm{K}})} \right) \ ,
\end{equation}
where $B_{\lambda}$ is the Planck function.  As was the case for direct imaging, 
the depths of absorption bands can be as large as the overall signature size (for 
strong features) or many times smaller (for weak features).

\section{Direct Detection of Surface Liquid Water}

Liquids, as opposed to diffusely scattering solid surfaces, have distinct polarization and 
scattering properties due to the process of Fresnel reflection \citep[][p.~382]{griffiths1999}. 
For a planar surface,  the polarization signature peaks at the Brewster angle, where the 
polarization fraction can approach unity for a liquid with no ripples or waves.  This surface 
will have enhanced reflectivity in the forward scattering direction where the observational 
angle of reflectance is equal to the solar angle of incidence (i.e.,~at the specular point), and 
this reflectivity increases towards glancing angles.

Measurements of the light polarization fraction may be an effective means to 
detect if an exoplanet has a surface ocean \citep{williams&gaidos2008,stam2008}.    
Earthshine and spacecraft observations reveal that Earth's polarization fraction is a 
function of the phase angle, peaking at values of 0.2--0.4 in the visible near quadrature 
\citep{coffeen1979}.  The location of this peak is crescent-ward of the maximum due to Rayleigh 
scattering, but depends on the wavelength-dependent competition between polarization 
from Rayleigh, cloud, haze, and ocean scattering \citep{zuggeretal2010,vaughanetal2023}.  Observing 
at near-infrared wavelengths will minimize the Rayleigh scattering contributions, pushing 
the polarization fraction peak to phase angles near those expected for ocean scattering 
(i.e., near 150$^{\circ}$), although peak polarization fractions are likely to still be 
$<0.2$ \citep{stam2008,zuggeretal2011,vaughanetal2023}.

\begin{figure}
\includegraphics[scale=.65]{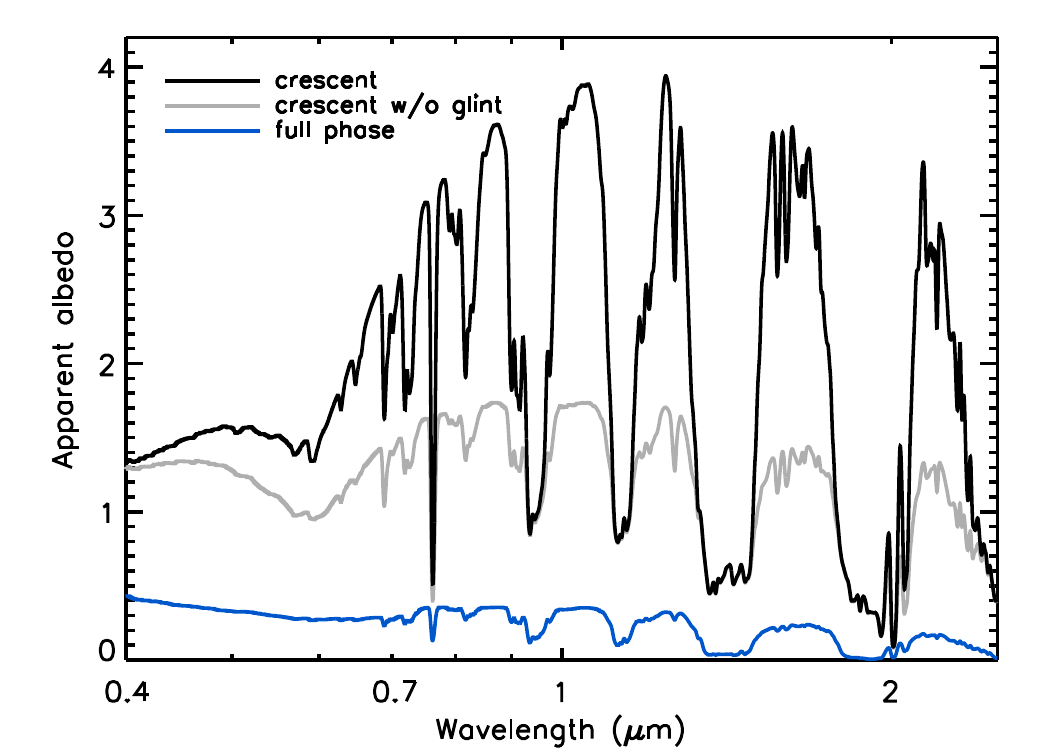}
\caption{Apparent albedo of Earth at full phase (blue), and at crescent phase both with 
              glint (black) and without glint (grey), from the validated model described in 
              \citet{robinsonetal2011}.  Apparent albedo is defined as the albedo a Lambert sphere 
              (with radius equal to the planetary radius) would need to reproduce the observed 
              brightness of the planet, and values larger than unity imply forward scattering.}
\label{fig:earth_alb}
\end{figure}

In addition to polarization, \citet{williams&gaidos2008} proposed that specular reflection 
from an ocean---which is often called ``glint''---could be used to detect surface liquid water 
on exoplanets.  Glint would manifest as an increasing planetary reflectivity towards crescent 
phase, and such an increase has been observed in Earthshine observations 
\citep{qiuetal2003,palleetal2003} and has been used to detect liquid seas in the polar regions 
of Titan \citep{stephanetal2010}.  Fig.~\ref{fig:earth_alb} shows apparent albedo spectra of 
Earth at full and crescent phases, including a crescent phase spectrum where glint is removed.

An increase in reflectivity towards crescent phase is not an unambiguous detection 
of glint, as forward scattering from clouds, Rayleigh scattering, and geometric effects 
can all produce a similar behavior.  \citet{robinsonetal2010} investigated 
the extent of the ocean glint effect using a model that included direction-dependent Rayleigh 
and cloud scattering.  This work showed that Rayleigh scattering false positives would be 
avoided by observing in the near-infrared, where the glinting Earth can be twice as bright as 
a non-glinting Earth.  Since snow/ice reflectivity decreases at longer wavelengths, observing in 
the near-infrared would also avoid the glint false positive discussed by \citet{cowanetal2012}, which 
explains a bias towards probing the icy polar regions of a planet at crescent phases.  Longitudinal maps 
made from rotationally-resolved photometry of an Earth-like exoplanet could reveal surface features whose 
reflectivity increases strongly at crescent phases \citep{lustigyaegeretal2018}, thereby providing 
another avenue to detect ocean glint effects and rule out surface false positives. The location 
of the maximum contribution from glint for Earth is near a phase angle of 150$^{\circ}$ and, for 
other planets, would depend on cloud cover, atmospheric thickness,  
and surface wind speeds.  As a proof of concept, \citet{robinsonetal2014b} were able to detect 
glint in unresolved {\it LCROSS} observations of Earth at a phase angle of 130$^{\circ}$, although 
this study benefited from significant {\it a priori} information.

\begin{figure}
\includegraphics[scale=.65]{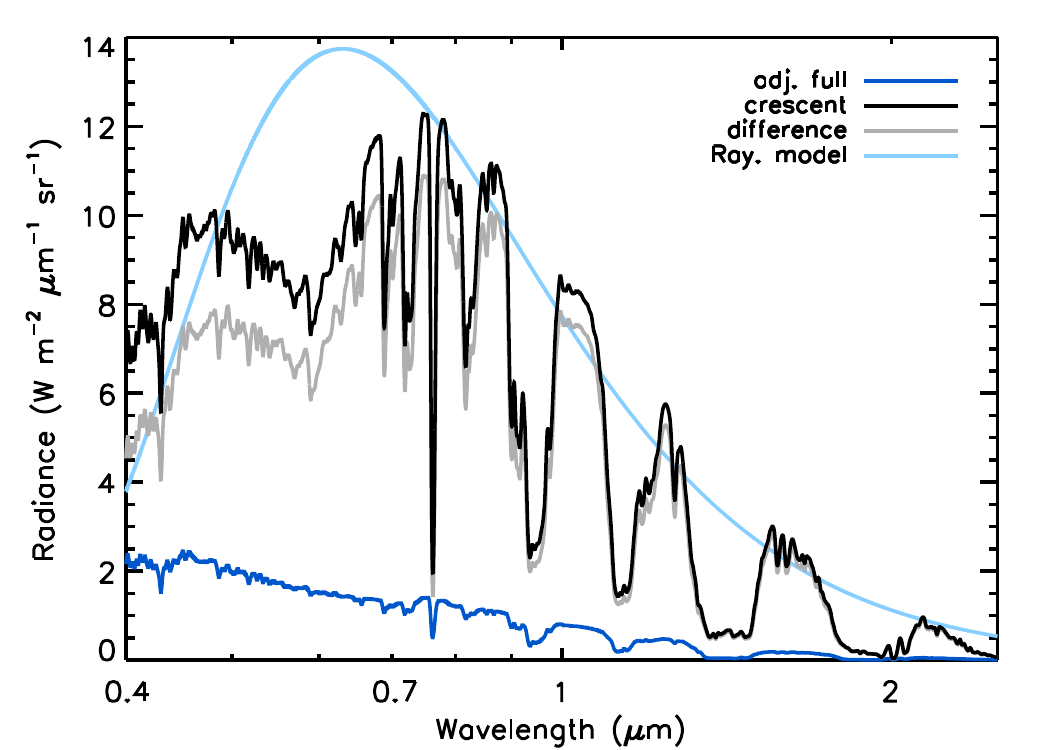}
\caption{Spectral signs of glint in phase-dependent observations of Earth.  A full phase spectrum of 
              Earth is corrected to crescent phase using a Lambert phase function (dark blue) and subtracted 
              from a crescent phase spectrum of Earth (black) to produce a difference spectrum (grey).  
              The result is well reproduced in window regions by a model that represents a solar spectrum 
              that experiences Rayleigh scattering opacity (light blue) as it passes twice through the 
              atmosphere and scatters once at the surface.}
\label{fig:earth_glint}
\end{figure}

Spectral information may prove essential for distinguishing between glint and its potential 
false positives.  \citet{robinson2012} noted that the unique atmospheric path traversed by a 
glint ray (i.e.,~two straight-line passes through the atmosphere with a single scattering event 
at the ocean surface) would imply that a significant portion of the crescent phase spectrum of 
Earth should resemble a solar spectrum modulated by Rayleigh scattering and gas absorption 
opacity.  Fig.~\ref{fig:earth_glint} demonstrates this signature.  Here, a full phase spectrum 
of Earth is corrected to crescent phase using a Lambert phase function, and is then subtracted 
from a crescent phase spectrum of Earth.  This difference spectrum, in essence, represents the 
forward scattering excess (due primarily to clouds and glint) at crescent phase.  In wavelength 
ranges with relatively little gas absorption and, thus, surface sensitivity, the difference spectrum 
can be well reproduced by a solar spectrum weighted by a term going as $\exp{(b/\lambda^4)}$ to 
account for Rayleigh scattering. As a demonstration of this effect, \citet{ryan&robinson2022} showed  
that the characteristic glint-spot spectrum could be detected using spectral principal component 
analysis on phase-dependent spectral models of Earth, given sufficient orbital access to 
crescent-phase spectroscopy or photometry.

Detecting glint or ocean polarization signatures requires, first and foremost, surface sensitivity and 
a favorable orbital inclination for the target so that the appropriate phase angles can be accessed.  
Measuring a polarization fraction of $f$ to a given signal-to-noise ratio (SNR) would require that the 
individual polarized flux measurements be improved by a factor of $\sqrt{2}/f$, implying the flux 
measurements be at a SNR of roughly 20 to detect $f\sim0.2$ at ${\rm{SNR}}=3$.  For the glint 
reflectivity and polarization signatures, which occur at somewhat extreme crescent phases, detection 
would require a minimum inclination of about 60$^{\circ}$. 
For such favorable inclinations, resolving an Earth twin at $2\lambda/D$ from a Sun-like host at 
10~pc would require an 8~meter diameter telescope (taking $\alpha = 150^{\circ}$ and 
$\lambda=1$~$\mu$m).  Notably, both the polarization and glint measurements can be accomplished 
with broadband observations, which helps to drive down requisite integration times.

\section{Surface Pressure and Temperature}

Numerous indicators impart information about pressure and temperature on spectra, although the 
scale of these signatures can sometimes be quite small.  The Rayleigh scattering optical depth is 
sensitive to the column abundances of primary atmospheric constituents, molecular lines (and 
bands) are broadened by pressure and thermal effects, and infrared spectra directly indicate the 
atmospheric and, possibly, surface thermal state.  Thus, depending on the wavelength range and 
observational approach, it is possible to use spectra to make inferences about the pressure and 
temperature at the surface of an exoplanet, thereby constraining habitability.

\begin{figure}
\includegraphics[scale=.60]{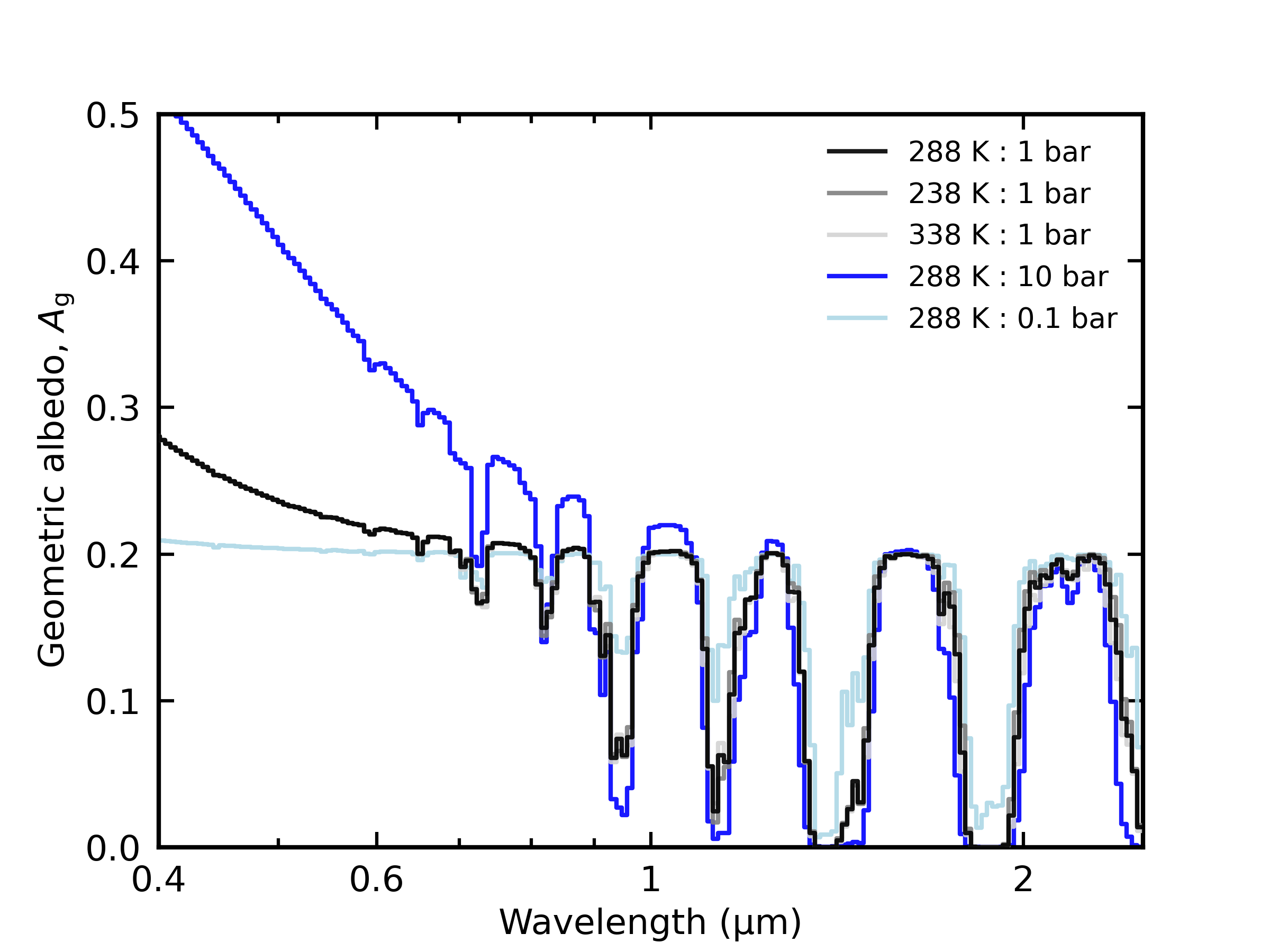}
\caption{Reflectivity spectra demonstrating pressure and temperature sensitivities.  All cases 
         are clearsky and have the same planetary mass and radius as Earth.  Water vapor is the only 
         absorbing gas and its mixing ratio profile is taken to be constant. The column mass 
         of water is fixed as Earth-like across all spectra.  The background gas is molecular 
         nitrogen.  Temperature effects are shown in grey colors (spanning 100~K) and pressure 
         effects are shown in blue colors (spanning two orders of magnitude).}
\label{fig:pT_sens}
\end{figure}

For reflected light observations, direct pressure and temperature information will come primarily 
from the Rayleigh scattering slope and from the widths of molecular bands.  Fig.~\ref{fig:pT_sens} 
shows reflectivity spectra of worlds with different atmospheric temperatures and pressures where 
water vapor is the only absorber.  The water vapor column mass is Earth-like and held fixed, so that 
variations are only due to pressure and temperature effects.  The influence of atmospheric 
temperature is rather limited.  Pressure, though, has a strong influence on band 
widths and depths, as well as the scale of the Rayleigh scattering slope. Raman scattering can also 
indicate atmospheric column density, is most apparent for surface pressures well in excess of 1 bar, 
and requires measurements at ultraviolet wavelengths \citep{oklopcicetal2016}

Our ability to extract pressure information from reflected light spectra will be complicated by 
uncertainties in planetary mass, planetary radius, the mass(es) of the primary atmospheric 
constituent(s), and clouds.  Nevertheless, inverse models applied to simulated \citep{fengetal2018} 
and real \citep{robinson&salvador2023} reflected light observations of the disk-integrated Earth 
indicate that spectral data with characteristic SNR of 10 can constrain surface pressure to within 
an order of magnitude and pushing the data quality to an SNR of 20 constrains the surface pressure 
to within a factor of 2–3. Complementary studies \citep{damiano&hu2022} with distinct treatments 
of planetary mass, clouds, and mixing ratio profiles for water vapor demonstrate that the pressure 
constraint is sensitive to the adopted spectral coverage for the exoplanet analog observation 
and is also likely sensitive to certain atmospheric parameterizations. Thus the connection 
between observational SNR and expected constraints on surface pressure remain unclear for 
Earth-like worlds studied in reflected light. Finally, limited study, outside of 
nitrogen- or hydrogen-dominated atmospheres, has been given to the dependence of pressure 
broadened line half widths on the composition of the background atmosphere 
\citep[e.g., ][]{gamacheetal1997,hedges&madhusudhan2016}.

Collision induced absorption features and dimer features---both of which are quite sensitive to 
background pressure---have been proposed as key indicators of pressure on exoplanets 
\citep{misraetal2014a}.  Such features have been detected in transmission spectra of Earth's 
atmosphere \citep{palleetal2009}, and in near-infrared spectra of the Pale Blue Dot 
\citep{schwietermanetal2015}.  In general, collision induced absorption and dimer features are  
usually stronger in reflected light and thermal emission spectra, since these techniques tend to 
probe deeper into atmospheres than transit spectra.  Strong molecular nitrogen pressure induced 
absorption is limited to a feature near 4.3~$\mu$m \citep{schwietermanetal2015}, key molecular 
oxygen features are at 1.06, 1.27, and 7~$\mu$m \citep{misraetal2014a}, carbon dioxide has a 
broad feature near 7~$\mu$m \citep{baranovetal2004}, and molecular hydrogen has numerous 
features spanning the visible, near-infrared, and thermal infrared 
\citep{frommholdetal2010,abeletal2011}.

Transit spectra, through their sensitivity to the atmospheric pressure scale height ($RT/\mu g$, 
where $R$ is the universal gas constant), contain additional information about temperature 
\citep{lecavelierdesetangsetal2008}, atmospheric mean molecular weight 
\citep{benneke&seager2012}, and, potentially, gravitational acceleration 
\citep[or mass;][]{dewit&seager2013} \citep[although see][]{batalhaetal2017}.  However, 
while a transit spectrum of an Earth twin would be rich with spectral features 
\citep{kaltenegger&traub2009}, there are a variety of processes that minimize (or prevent) 
sensitivity to the surface environment.  Fundamental amongst these processes is atmospheric 
refraction, where rays passing through deeper regions of a planetary atmosphere can experience 
enough refraction to bend them off the stellar disk 
\citep{betremieux&kaltenegger2014,misraetal2014b}.  For Earth-like planets around 
Sun-like stars, refraction prevents sensitivity to the troposphere---although a small amount 
of the light that passes through the planet's atmosphere near transit ingress and egress can 
follow paths that pass near the surface \citep{misraetal2014b}.  

\begin{figure}
\begin{tabular}{c}
\hspace{2cm}
\includegraphics[scale=.50]{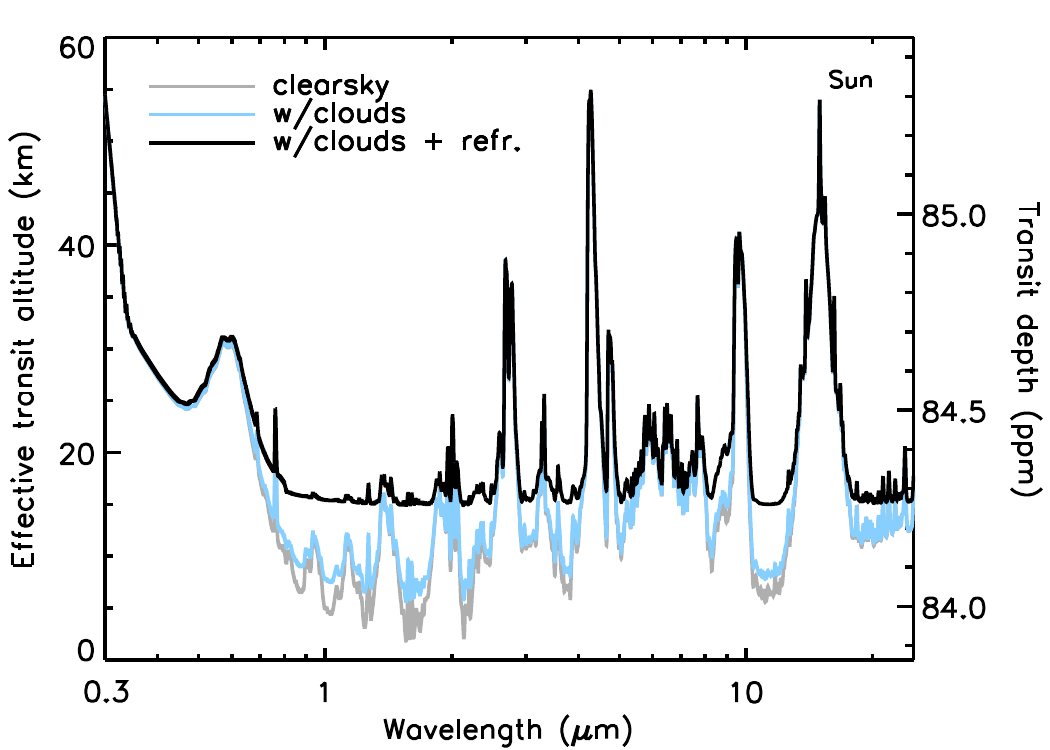} \\
\hspace{2cm}
\includegraphics[scale=.50]{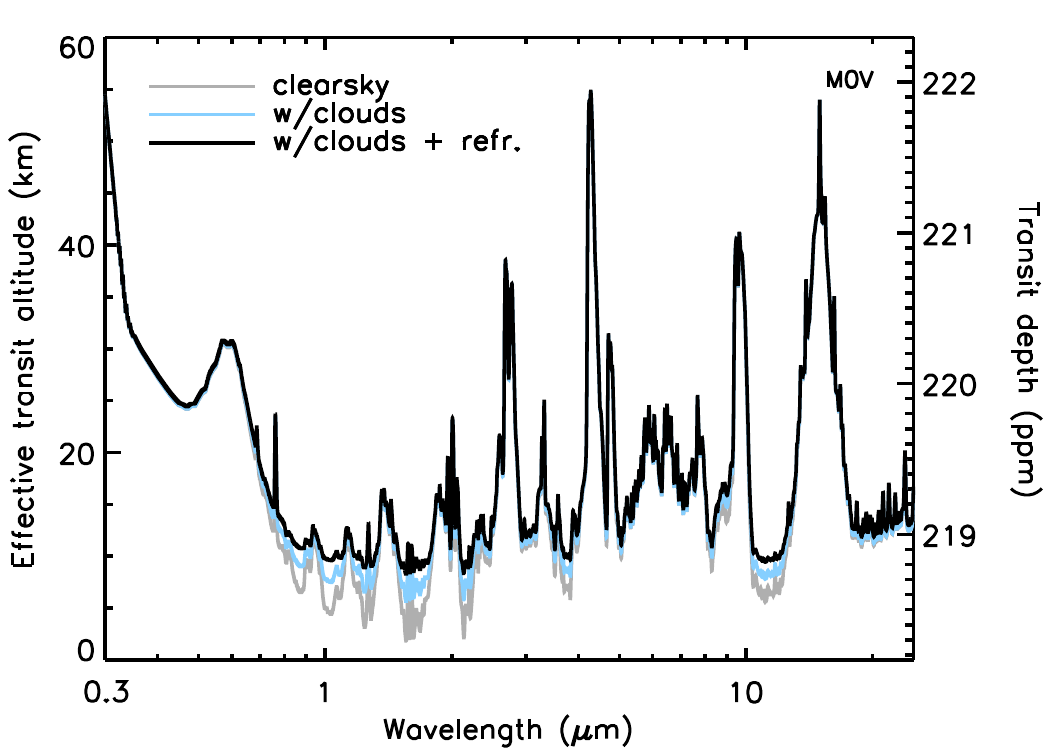} \\
\hspace{2cm}
\includegraphics[scale=.50]{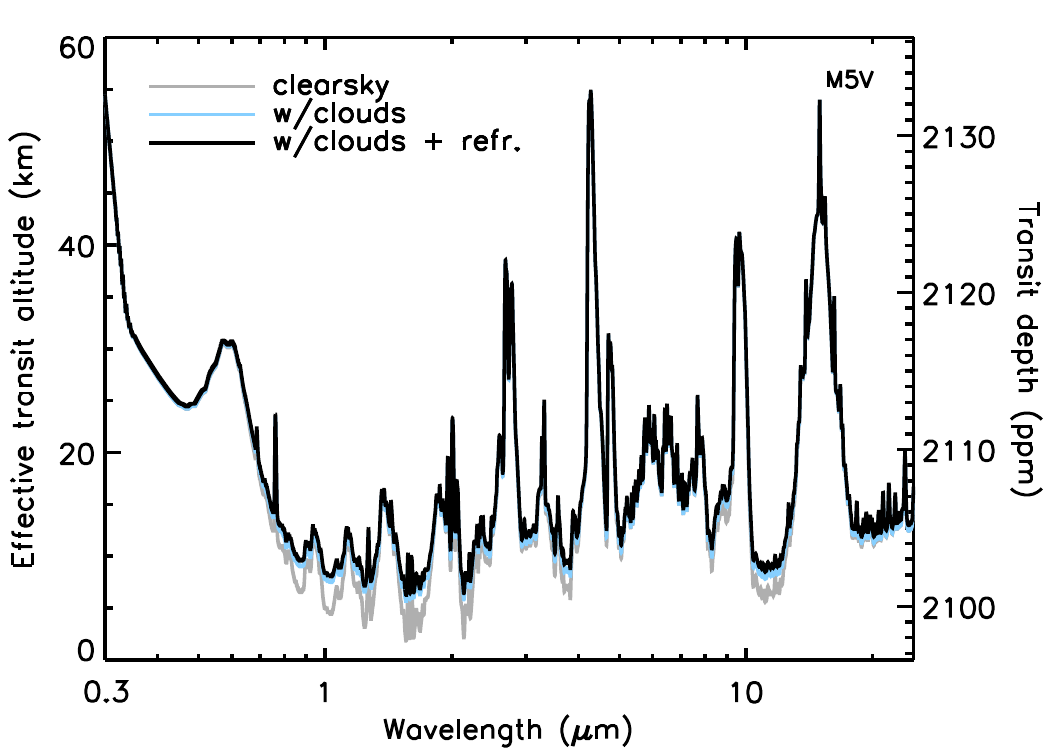} \\
\end{tabular}
\caption{Transit spectra of an Earth twin around a Sun-like, M0 dwarf, and M5 dwarf host.  Clearsky 
              cases are in grey, blue curves show the addition of realistic clouds, and black curves contain 
              both clouds and refraction. From a transit spectra model described in \citet{robinson2017}.}
\label{fig:trans}
\end{figure}

Transit spectra of  Earth twins around mid- or late-M dwarfs are much less affected by 
refraction, so, here, clouds and gas opacity are the primary impediment to surface sensitivity.
Large slant pathlengths associated with the transit geometry cause 
spectral regions with only a small amount of vertical optical depth to become opaque.  This 
issue is especially true for clouds and hazes \citep{fortney2005}, which tend to have optical 
depths that vary slowly in wavelength, and can thus block observations of the deep 
atmosphere over wide spectral ranges 
\citep[e.g.,][]{kreidbergetal2014,knutsonetal2014,robinsonetal2014a}.  Aerosol extinction 
in transit spectra would be significant for a hazy early Earth \citep{arneyetal2016}, and even 
modern Earth, which is generally thought to have ``patchy'' clouds, is $>\!\!70$\% cloud-covered 
when thin, high-altitude cirrus clouds are considered \citep{stubenrauchetal2013}.  
Fig.~\ref{fig:trans} shows transit spectra of Earths around different host star types, and 
includes the effects of realistic clouds and refraction.

Achieving the types of transit observations outlined in this section is made difficult by detector 
noise and systematics and clouds.  \citet{greeneetal2016} showed that a single visit with {\it James Webb Space 
Telescope} ({\it JWST}), with an assumed set of instrument systematic noise floors, is unlikely to be 
able to place strong constraints on the atmospheric properties of a warm (500~K) super-Earth planet 
with a steam atmospheres transiting an M0 dwarf.  By pushing to 
a mid-M dwarf host, \citet{benneke&seager2012} concluded that {\it JWST} could place constraints on 
the surface pressure of a similarly warm super-Earth planet with an atmosphere dominated by molecular 
nitrogen.  The \citet{benneke&seager2012} results did not address clouds, refraction, detector 
systematics, or cooler (Earth-like) atmospheres, all of which will make detections of surface pressure 
more difficult.  Most promisingly, \citet{dewit&seager2013} found that {\it JWST} could constrain 
pressures and temperatures for a cloud-free Earth-like planet orbiting a late-M dwarf using 200~hr of 
in-transit observation time, although potential degeneracies may exist \citep{batalhaetal2017}. Early 
transit observations of Earth-sized worlds orbiting late M-type stars with JWST have not yet yielded 
atmospheric constraints and indicate that substantial investments of observing time will be 
required to obtain effective reconnaissance of these atmospheres \citep{limetal2023}.

A limited number of studies exist that address the information content (especially with regard 
to surface pressure and temperature) of directly observed emission spectra from Earthlike 
exoplanets. Using retrieval techniques to fit simulated observations of a cloud-free Earth in 
the thermal infrared,  \citet{vonparisetal2013} showed that surface temperatures and pressures 
for Earth-like planets could be constrained to within the habitable range with spectral resolving power  
($\mathcal{R}=\lambda/\Delta \lambda$) greater than roughly 10 and SNRs greater than roughly 10, 
although these estimates are likely optimistic given the cloud-free assumption made in this work.  
More-recent cloud-free inverse modeling studies in support of the Large Interferometer For 
Exoplanets (LIFE) mission concept\,---\,a space-based mid-infrared interferometer capable of imaging 
and characterizing rocky worlds in the habitable zone of nearby stars \citep{quanzetal2022}\,---\,indicate 
that larger SNRs (greater than roughly 20) might be required to constrain surface temperatures to be 
above the water freezing point \citep{aleietal2022}. Retrieval studies applied in the thermal infrared 
to Earth-like worlds that do include clouds tend to agree with the requirement of SNRs larger than 
roughly 20 for strong surface temperature constraints \citep{robinson&salvador2023}.

\section{Other Habitability Indicators}

A variety of observations, while not direct confirmations of habitability, could also be used 
as evidence for liquid water at/near the surface of an exoplanet.  Within this area, the topic that 
has seen the most study is that of photometric variability.  At visible wavelengths, contrast 
between Earth's reflective clouds and its surface---which is absorptive due to the large ocean  
coverage fraction---makes our planet the most variable in the Solar System 
\citep{fordetal2001}, with peak-to-trough diurnal variations typically of order 20\% 
\citep{livengoodetal2011}.

Rotationally resolved, visible wavelength observations of the Pale Blue Dot could be used to 
produce surface feature maps \citep{cowanetal2009,fujiietal2011,cowanetal2011}, and lightcurves 
resolved over longer timescales could indicate variability due to weather.  (Lightcurves at thermal 
wavelengths could also reveal variability due to weather, but have received little study.)  Similar 
observations for a distant Earth-like planet, coupled with information about the planetary orbit (or 
insolation), would likely argue for atmospheric water vapor condensation, although the condensate 
phase (liquid or solid) and whether or not the aerosols reach a surface in a liquid state would be 
difficult to discern.  Confirmation, or detection, of the presence of liquid droplets, as well as their 
composition, could come from reflectance and polarization measurements at phase angles 
corresponding to maximum scattering from the primary rainbow of the droplets 
\citep{bailey2007,vaughanetal2023}, although accessing these phase angles requires orbital 
inclinations like those needed for glint measurements.

Detecting other signs of water vapor condensation, especially near an exoplanetary surface, 
would make for stronger indications of habitability.  \citet{fujiietal2013} used rotationally resolved 
spectra of the Pale Blue Dot to detect differences in the spatial distribution of water vapor and 
molecular oxygen in Earth's atmosphere.  Since molecular oxygen is well-mixed, this detection 
argues for a non-uniform vertical and horizontal distribution of water vapor, where the most 
likely interpretation for a potentially habitable exoplanet would be exchange between the gas 
and liquid/solid phase.  More recently, \citet{robinson&marley2016} noted that retrieval of a 
water vapor mixing ratio profile that is larger near an exoplanetary surface, or the retrieval of a 
condensate cloud layer located near the surface of a potentially habitable exoplanet, would 
argue for stable surface liquid or solid water. Recent inverse modeling applied to synthetic 
reflected light observations of Earthlike worlds reveals that detecting signs of structure in 
a water vapor profile will be extremely challenging, possibly requiring SNRs larger than 20 
for both optical and near-infrared coverage \citep{damiano&hu2022}.

\begin{figure}
\includegraphics[scale=.65]{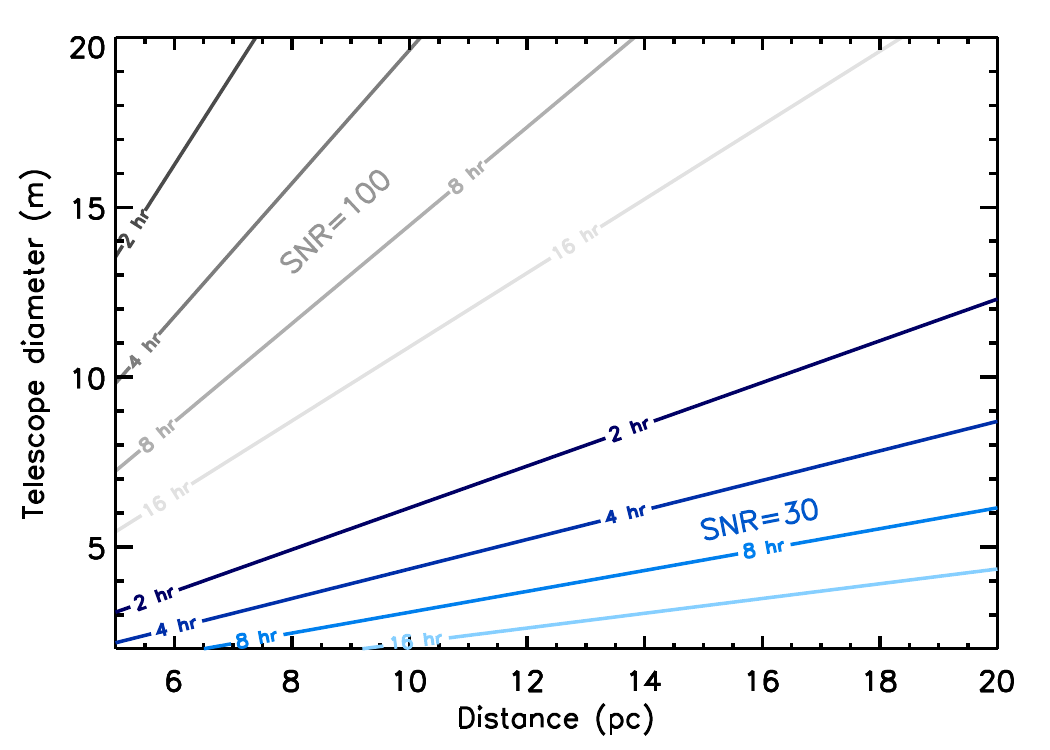}
\caption{Contours of integration time required to achieve a SNR of 100 (grey) and 30 (blue) 
               for Earth twins in V band as a function of telescope diameter and distance to the 
               planetary system.  Only noise from stellar leakage (at a raw contrast of 
               $10^{-10}$), Solar System zodiacal light, and exozodiacal light (at the level of 
               three exozodis) are considered.  Models assume a Sun-like host, and relevant 
               expressions are given in \citet{robinsonetal2016}.  Achieving a SNR of 100 would 
               be strongly limited by systematic noise floors.}
\label{fig:earth_Dt}
\end{figure}

Mapping of Pale Blue Dots is made difficult by the requisite SNRs, and, potentially, the need 
for simultaneous photometry in multiple bands (although these bands can be wide, which helps 
to increase the signal from the planet).  \citet{cowan&strait2013} use diurnal lightcurves at a 
SNR of 100 to map the Pale Blue Dot without prior information for surface or cloud spectra, 
although simply detecting changing cloud patterns may only require SNRs larger than roughly 
20--30.  Fig.~\ref{fig:earth_Dt} shows how the integration time required for V band photometry 
depends on distance to the system and telescope diameter.  If SNRs of 100 are required within 
2--4 hr (for rotational resolution), then mapping will be limited to only very nearby targets.  
However, if mapping can be accomplished with lower SNRs, then targets out to much larger 
distances can be accessed, even with modest-sized telescopes.  Finally, note that retrieving 
water vapor or cloud profiles from reflected light observations will also require high-SNR 
observations, at least if the jovian cloud retrievals mentioned in the previous section are indicative.

\section{Outstanding Challenges}

\begin{table}
\caption{Key Habitability Observables and Constraints for Earth Twins.}
\label{tab:obs} 
\begin{tabular}{p{2.3cm}p{2.0cm}p{1.5cm}p{2.0cm}p{3.3cm}}
\hline\noalign{\smallskip}
Observable & Technique & Wavelength & Noise Req.  & Add'l Considerations \\
\noalign{\smallskip}\svhline\noalign{\smallskip}
glint                                    & direct imaging  & 0.7--2.5 $\mu$m & ${\rm{SNR}} \gtrsim 3$        & broadband; $i \gtrsim 60^{\circ}$ \\[6pt]
\hline
polarization                        & direct imaging & 0.7--2.5 $\mu$m & ${\rm{SNR}} \gtrsim 20$      & broadband; $i \gtrsim 10^{\circ}$ \\[6pt]
\hline
                                          & transit              & 0.4--30 $\mu$m  & $\lesssim 10$--50 ppm          & $\mathcal{R} \gtrsim 100$; mid/late-M dwarf \\[6pt]
surface $p$ \& $T$            & direct imaging  & 0.4--2.5 $\mu$m & ${\rm{SNR}} \gtrsim 20$ (?)      & $\mathcal{R} \gtrsim 100$; no $T$ constraint \\[6pt]
                                           & direct imaging  & 4--30  $\mu$m    & ${\rm{SNR}} \gtrsim 20$ &  $\mathcal{R} \gtrsim 10$ \\[6pt]
\hline
weather/mapping               & direct imaging & 0.4--1 $\mu$m & ${\rm{SNR}} \gtrsim 30$--100 & broadband \\[6pt]
\hline
H$_{2}$O/cloud profiles    & direct imaging  & 0.4--2.5 $\mu$m & ${\rm{SNR}} \gtrsim 20$ (?)      & $\mathcal{R} \gtrsim 100$ \\[6pt]
\noalign{\smallskip}\hline\noalign{\smallskip}
\end{tabular}
\end{table}

While a variety of techniques and observables relevant to characterizing habitability have 
been proposed, key questions still remain about the feasibility and utility of these different 
methods.  Regarding the direct detection of surface liquid water, requisite integration 
times for realistic observing scenarios have seldom been explored 
\citep[although see][]{lustigyaegeretal2018,ryan&robinson2022}.  Performing these observations 
in the near-infrared (where stars are fainter) may prove costly, and noise from observing 
near the inner working angle (for glint) or from polarized light from exozodiacal dust will both 
introduce complications.  Similarly, SNRs (which dictate integration times) required for 
retrieving water vapor and cloud profiles from reflection, emission, and/or transmission 
spectra of realistic Pale Blue Dots also remain largely unexplored.  
Finally, future work on variability should focus on the wavelength range, timing, and minimum 
required SNRs to do mapping.  Table~\ref{tab:obs} presents an overview of the current 
understanding of the observing requirements for the different approaches to detecting or 
constraining habitability.

Once observational feasibility has been addressed, a more holistic discussion of 
characterizing for habitability should emerge.  This will be especially true for high 
contrast imaging.  Here, repeat observations may be required to confirm the planetary 
nature of a target and to constrain the orbit (and, thus, insolation) of a confirmed planet.  
It is unclear if certain observational tests for habitability should be worked into the 
confirmation and orbit determination sequence.  Also, following this sequence, open 
questions remain regarding the order in which different observations (e.g., glint, polarization, 
moderate resolution spectroscopy) should take place.  Such questions can only be 
settled by weighing the information supplied by these different observations with the time 
required to achieve them.

\section{Conclusions}

Detecting or constraining the habitability of a distant exoplanet will be a challenging and 
critical step towards understanding the frequency of the origin of life on other worlds.  
Transit spectroscopy, secondary eclipse observations, and high contrast imaging all 
have the potential to reveal key planetary properties related to habitability, and these 
techniques each have their own assets and challenges.  Reflected light observations 
can directly reveal surface liquid water, either through polarization or glint measurements.  
Constraints on surface pressure are possible with most observational techniques (depending 
on wavelength coverage), but surface temperatures (which, when combined with a surface 
pressure measurement, can demonstrate habitability) will prove difficult to measure in 
reflected light.  Detecting water vapor condensation at/near a surface is also feasible, either 
through spectral retrieval of gas mixing ratio or condensate profiles, or through mapping 
using time resolved photometric measurements.  In the end, though, it could prove that no 
unambiguous ``smoking gun'' exists for detecting stable surface liquid water, so that actual 
constraints on habitability may come from multiple lines of evidence using a variety of 
approaches brought to bear on a distant Pale Blue Dot.

\begin{acknowledgement}
TR gratefully acknowledges support from NASA through the Sagan Fellowship 
Program executed by the NASA Exoplanet Science Institute. The results reported 
herein benefitted from collaborations and/or information exchange within NASA's 
Nexus for Exoplanet System Science (NExSS) research coordination network 
sponsored by NASA's Science Mission Directorate.  Certain essential tools used 
in this work were developed by the NASA Astrobiology Institute's Virtual Planetary 
Laboratory, supported by NASA under Cooperative Agreement No. NNA13AA93A.
TR thanks J Fortney for a constructive critique of this review.
\end{acknowledgement}



\end{document}